\documentclass{aastex}
\usepackage{spr-astr-addons}
\usepackage{url}
\usepackage{graphicx}

\RequirePackage{color}

\def\simlt{\ \raise -2.truept\hbox{\rlap{\hbox{$\sim$}}\raise5.truept   %
\hbox{$<$}\ }}
\def\simgt{\ \raise -2.truept\hbox{\rlap{\hbox{$\sim$}}\raise5.truept   %
\hbox{$>$}\ }}

\begin{document}

\title{Probing the low-energy spectrum of non-thermal electrons in galaxy clusters with soft gamma ray observations}
\slugcomment{}
\shorttitle{Low-energy electrons in galaxy clusters}
\shortauthors{P. Marchegiani}

\author{P. Marchegiani\altaffilmark{1}} 
\altaffiltext{1}{School of Physics, University of the Witwatersrand, Private Bag 3, 2050-Johannesburg, South Africa. Email: Paolo.Marchegiani@wits.ac.za}

\begin{abstract}
In this paper we study the possibility of probing the low-energy part of the spectrum of non-thermal electrons in galaxy clusters by detecting their non-thermal bremsstrahlung (NTB) emission in the soft gamma ray band, using instruments like e-ASTROGAM. Using the Coma cluster as a reference case, we find that, for very low values of the minimum energy of the electrons, in principle the NTB is detectable, but this situation is possible only for conditions that can be maintained only for a short time compared to the cluster lifetime. The possibility of constraining the low energy spectrum of non-thermal electrons through NTB is therefore hard to achieve in next years. 
\end{abstract}


\section{Introduction}

Our knowledge of the properties of non-thermal electrons in galaxy clusters is based on the observation of diffuse radio emission like radio halos and relics, that are produced by synchrotron emission of relativistic electrons having energy of the order of GeV interacting with magnetic fields with intensity of the order of $\mu$G (e.g. Feretti et al., 2012). The same electrons are also expected to produce Inverse Compton Scattering (ICS) emission when interacting with the Cosmic Microwave Background (CMB) photons (Perola \& Reinhardt, 1972); however, this emission has not yet been firmly detected even in nearby and rich clusters like Coma (e.g. Gastaldello et al., 2015).

While there are hints of a steepening of the electrons spectrum at high energies giving origin to an analogous steepening in the integrated radio halo spectrum (see Thierbach et al., 2003 for the Coma case), the lower energy part of the electrons spectrum remains basically unconstrained, with the possible exception of the detection in a few clusters of an excess in the Extreme Ultraviolet band (EUV; see e.g. Lieu et al., 1996, Bowyer et al., 2004), that can be due to ICS of electrons with energy of the order of 100--200 MeV (Sarazin \& Lieu, 1998), but also to thermal bremsstrahlung from gas with low temperature (Lieu et al., 1996).

Due to the steep spectra usually found for non-thermal electrons in galaxy clusters (see e.g. Feretti et al., 2012), the low-energy part of the electrons spectrum is expected to provide the largest contribution to the total number and the total energy content of these electrons. Therefore it would be important to identify a way to probe the electrons spectrum at low energies, in order to estimate the ratio between the energy content stored in non-thermal electrons and the other components of the Intra Cluster Medium (ICM), like thermal gas, magnetic field, and non-thermal protons (see, e.g., Colafrancesco \& Marchegiani, 2011 for an analogous discussion in radio galaxies).

Electrons with energies lower than $\sim100$ MeV are expected to have a short lifetime compared with the cluster one because of the Coulomb interactions with the thermal electrons, with the lifetime being shorter for lower electrons energies (e.g. Sarazin, 1999). As a results, non-thermal electrons spectra that are injected with an initial power-law shape are expected to flatten at low energies (e.g. Lieu et al., 1999); this fact can be phenomenologically parametrized through the value of the minimum Lorentz factor $\gamma_{min}$ for which the electrons spectrum can be described by a power law for $\gamma>\gamma_{min}$ (e.g. Colafrancesco \& Marchegiani, 2011). The energy content stored in the non-thermal electrons is strongly depending on this value.

A possible way to probe the low-energy part of non-thermal electrons spectrum in galaxy clusters is given by the study of the non-thermal Sunyaev-Zel'dovich effect (SZE), i.e. the distortion of the CMB spectrum by ICS produced in ionized media present in cosmic structures (e.g. Colafrancesco, Marchegiani \& Palladino, 2003). A non-thermal contribution to the total SZE has been possibly detected in the Bullet cluster (Colafrancesco, Marchegiani \& Buonanno, 2011; Marchegiani \& Colafrancesco, 2015), suggesting the possibility of a very small value of the minimum energy of the electrons ($\gamma_{min} \sim 1.4$). Also, a statistical study of the SZE in 23 radio halo clusters (Colafrancesco et al., 2014) suggested an average value of the ratio between non-thermal and thermal electrons pressures of the order of $X\equiv P_{non-th}/P_{th}\approx 0.55$. Moreover, an interpretation of a stacked analysis of Planck HFI data in galaxy clusters (Hurier, 2016) in terms of a non-thermal contribution to the SZE (Marchegiani \& Colafrancesco, 2017) has indicated a low value of $\gamma_{min}$ in this sample, that corresponds to a possible pressure ratio of the order of $X\sim20-30\%$. These results are in contradiction with the values of the upper limits on the pressure ratio between non-thermal protons and thermal gas derived from Fermi-LAT gamma ray upper limits in galaxy clusters, that are of the order of $2-6\%$ (Huber et al., 2013, Prokhorov \& Churazov, 2014); therefore, it would be useful to identify another possible way to detect low-energy non-thermal electrons, in order to check the validity of these results.

Non-thermal electrons with $E<100$ MeV should emit by non-thermal bremsstrahlung (NTB) when interacting with the thermal gas nuclei (e.g., En\ss lin, Lieu \& Biermann, 1999); this radiation should be emitted in the region of the electromagnetic spectrum under 100 MeV, that is presently almost unexplored, but can become accessibile in the future thanks to proposed instruments like e-ASTROGAM\footnote{http://eastrogam.iaps.inaf.it/} and AMEGO\footnote{https://asd.gsfc.nasa.gov/amego/}. Therefore in this paper we study the possibility to use NTB as a probe of low-energy non-thermal electrons at the light of the properties of these instruments, using the Coma cluster as a case of study.
Coma is a rich and nearby cluster where a bright radio halo has been observed (e.g. Deiss et al., 1997), meaning that a high level of diffuse non-thermal activity is present; for these reasons, it looks as one of the candidates where in principle the observation of non-thermal emissions is more favored compared to other clusters.

We adopt in this paper a simple phenomenological model, i.e. a power-law spectrum for non-thermal electrons for $\gamma>\gamma_{min}$, and explore the consequences on the produced NTB by varying the values of $\gamma_{min}$. Since the angular resolution of the proposed instruments (de Angelis et al., 2018) is larger than the Coma cluster angular size, we consider the emission integrated on the whole cluster as a point-like source.

The outline of the paper is the following: in Section 2 we describe the theoretical methods used in this paper. In Section 3 we present the results, and summarize and discuss our results in Section 4. Throughout the paper, we use a flat, vacuum--dominated cosmological model following the results of Planck, with $\Omega_m = 0.308$, $\Omega_{\Lambda} = 0.692$ and $H_0 =67.8$ km s$^{-1}$ Mpc$^{-1}$ (Ade et al., 2016).

\section{Methods}

NTB is produced by interactions of non-thermal electrons with thermal nuclei. For an electrons spectrum giving the particle density per unit of energy and volume $N_e(E,r)$, the NTB emissivity is given by: 
\begin{equation}
j_{B}(E_\gamma,r)=\int d E N_e(E,r) P_{B}(E_\gamma,E,r)
\end{equation}
(e.g. Schlickeiser, 2002), where $E_\gamma$ is the energy of the emitted photon, and
\begin{equation}
P_B(E_\gamma,E,r)=\beta c E_\gamma   n_{th}(r) \sigma(E_\gamma,E),
\label{eq.pb.brem}
\end{equation}
where $n_{th}(r)$ is the thermal gas density, and where the cross section for relativistic electrons is given by:
\begin{eqnarray}
\sigma(E_\gamma,E) & = & \frac{1}{E_\gamma} \frac{3}{8\pi} \alpha \sigma_T \left[ \left[ 1+
\left(1-\frac{E_\gamma}{E} \right)^2 \right] \phi_1 + \right. \nonumber \\
& & - \left. \frac{2}{3}
\left(1-\frac{E_\gamma}{E} \right) \phi_2 \right],
\end{eqnarray}
being $\alpha=1/137.036$ the fine-structure constant and $\sigma_T=6.652\times10^{-25}$ cm$^{2}$ the Thomson cross section. For unshielded target nuclei (like in the case of the ionized intra cluster medium) we can assume $\phi_1=\phi_2=Z^2\phi_u$, with $Z=1$ and
\begin{equation}
\phi_u=4 \left[ \ln \left[ \frac{2E}{m_e c^2} \left( \frac{E-E_\gamma}{E_\gamma}
\right) \right] - \frac{1}{2} \right].
\end{equation}
In the following we assume for the non-thermal electrons spectrum a simplified phenomenological model, assuming that they have a radial profile proportional to the thermal gas one $g_{th}(r)$, and a power-law spectrum with a low-energy cut-off: 
\begin{equation}
N_e(\gamma,r)=k_0 \gamma^{-s_e}g_{th}(r) \;\; \mbox{for } \gamma \geq \gamma_{min}  ,
\label{spettroele}
\end{equation}
with a typical value of the spectral index in galaxy clusters $s_e=3.7$, and where the parameter $k_0$ provides the normalization of the non-thermal electrons density. 

The thermal gas density profile in the Coma cluster is derived from X-ray measures:
\begin{equation}
n_{th}(r)=n_0 g_{th}(r) = n_0 \left[1+\left(\frac{r}{r_c}\right)^2\right]^{-q_{th}}
\end{equation}
with parameters  $n_0=3.4 \times 10^{-3} h_{70}^{1/2}$ cm$^{-3}$, $r_c= 300 h_{70}^{-1}$ kpc and $q_{th}=1.125$ taken from Briel, Henry \& B\"ohringer (1992).

As effect of NTB, it is expected that an electron with energy $E$ has the maximum of its emission at approximately $E_\gamma\sim E/2$, and that, for an electrons spectrum as in eq.(\ref{spettroele}), the NTB has a flux spectrum $F_B\propto E_\gamma^{-s_e+1}$ (e.g. Longair, 1994). As a consequence, we expect the NTB has a steep photon spectrum, with a sudden low-energy break at an energy $E_{\gamma}\sim \gamma_{min} m_e c^2/2$, under which the photons spectrum should strongly decrease.

Non-thermal electrons emit also by ICS with the CMB photons in a frequency range going from EUV to gamma rays, if the electrons spectrum does not steepen at high energies, with a spectral index $\alpha_{ICS}=-(s_e-1)/2$, i.e. flatter than the NTB one. The combination of NTB and ICS therefore has a flatter component, given by ICS, with a peak produced by NTB at an energy around $E_{\gamma}\sim \gamma_{min} m_e c^2/2$; the detection of such a peak can in principle allow to derive the value of $\gamma_{min}$.

Non-thermal electrons emit also by synchrotron in the radio band when interacting with the intra-cluster magnetic field. The intensity and the spatial profile of the magnetic field can be derived from Faraday Rotation measures, that in the case of Coma provide as a best fit a central value of $B_0=4.7$ $\mu$G and a radial profile proportional to $g_{th}^{1/2}(r)$ (Bonafede et al., 2010). In the following we normalize the electrons density $k_0$ by requiring that the radio flux produced at 1.4 GHz for these properties of the magnetic field is equal to the observed value of 640 mJy (Deiss et al., 1997); we choose this frequency because it is the highest frequency where the spectrum of the Coma radio halo has a power-law shape (Thierbach et al., 2003). We also note that this simple model does not take into account the possible effect of re-acceleration of electrons that can be provided by turbulences, that can boost the electrons spectrum only for electrons that emit in the radio band, i.e. with $\gamma\sim10^3-10^4$ (e.g. Brunetti et al., 2001). As a consequence, in presence of this boosting produced by turbulences, the ratio between the number of electrons emitting in radio and the ones with lower energies would be higher, and the flux produced by NTB would be reduced compared to the radio one. For this reason, in our simplified model the resulting NTB should be considered as an upper limit. 

The low-part of the electrons spectrum is basically unconstrained; at low energies the electrons loose their energy mainly because of the Coulomb interactions with the thermal gas, and for low energies the electrons lifetime can be smaller than $10^8$ yrs (e.g. Sarazin, 1999); low energy electrons therefore should be present only for a short time after being accelerated. Electrons with $\gamma\sim100-200$ should instead have a longer lifetime, even comparable with the cluster one (e.g. En\ss lin et al., 1999, Brunetti et al., 2001), so it is reasonable to think that a low-energy cutoff in the electrons spectrum should not be higher than these values.

In the following, we normalize the density of non-thermal electrons to the radio flux at 1.4 GHz as described, and for several values of $\gamma_{min}$ we calculate the ICS and NTB emissions in the soft gamma ray band, comparing the results with the expected sensitivity of e-ASTROGAM; we also calculate the energy content stored in non-thermal electrons compared to the thermal one, and the heating rate provided by non-thermal electrons compared with the cooling rate due to thermal bremsstrahlung, in order to check if the physical configurations associated to these assumptions are problematic for the stability of the ICM.

The energy content stored in the non-thermal electrons is given by:
\begin{equation} 
U_e  =  \int_{\gamma_{min}}^\infty d\gamma N_e(\gamma) (\gamma-1) m_e c^2  ,
\end{equation}
that can be compared with the energy stored in the thermal gas, $U_{th}=3n_{th}k_B T_e$, that is $\sim1.3\times10^{-10}$ erg cm$^{-3}$ at the center of the Coma cluster.

Non-thermal electrons can also heat the ICM because of Coulomb interactions. The heating rate induced by the electrons having the spectrum assumed in Eq.(\ref{spettroele}) is given by:
\begin{equation}
\dot{ \epsilon}_h \equiv {d \epsilon \over dt} \Biggr|_{_{\rm heat}} = \int_{E_{min}}^{\infty } N_{e}(E)
 \bigg({dE \over dt}\bigg) dE \; .
 \label{eq.heating}
\end{equation}
For a relativistic electron with velocity $v=\beta c$ and Lorentz factor $\gamma$, the heating rate on thermal gas due to Coulomb interactions is:
\begin{equation}
{dE\over dt}\approx K \, Z^2\,{1\over \beta}\,
   \left [\ln {2\, m_{\rm e}\, c^2\, \beta^2 \gamma^2\over I_{\rm p}}
 -\beta^2\right],
\label{dedx}
\end{equation}
where $Z^2$ is the (suitably averaged) squared charge of the plasma's nuclei, $K\!=\!4\,
\pi\, n_{\rm th}\, r_{\rm e}^2\, m_{\rm e}\, c^3$ and $r_{\rm e}\!=\!e^2/m_{\rm e}\,
c^2\!\simeq\!2.82$ fm.
Here $I_{\rm p}\!=\!\hbar\, \omega_{\rm p}$, with $\omega_{\rm p}\!=\! [4\pi\,
n_{\rm th}\,e^2/m_{\rm e}]^{1/2}$ being the plasma frequency (e.g. Colafrancesco \& Marchegiani, 2008).
The heating rate can be compared with the cooling rate of the thermal gas, given by
\begin{equation}
\dot{ \epsilon}_c \equiv {d\epsilon\over dt}\Biggr|_{_{\rm cool}}=  \sqrt{2^{11}\pi^3\over 3^3}\; {e^6\sqrt{m_{\rm e} }\over h\,m_{\rm e}^2\, c^3}\,
{\bar G}\, {\bar z} \, n_{\rm th}^2\, \sqrt{kT} 
\end{equation}
(e.g. Longair, 1994), where $\bar z$ is an average charge of the IC plasma and ${\bar G}$ is the Gaunt factor; the cooling rate at the center of the Coma cluster is $\dot{ \epsilon}_c\sim1.6\times10^{-28}$ erg cm$^{-3}$ s$^{-1}$. The values of the heating rate are dependent on the value of the normalization of the electrons spectrum and the electrons minimum energy (see eq.\ref{eq.heating}), and for the models we are using are reported in Table \ref{tab1}.

\section{Results}

The normalization of the electrons spectrum required to not produce a radio emission in excess compared to the observed one is $k_0=1.6\times10^{-4}$ cm$^{-3}$; since the radio emission is not sensible to the low-energy spectrum of electrons, this value does not depend on $\gamma_{min}$. We consider several value of $\gamma_{min}$ between 3 and 200, and calculate the energy content stored in the relativistic electrons and the heating rate provided by them. The results are in Table \ref{tab1}.

\begin{table}[t]
\begin{center}
 \caption{Values of $\gamma_{min}$ and corresponding values of the ratio of the energy content stored in non-thermal electrons and in the thermal gas, of the heating rate, and of the ratio between heating and cooling rates.}
 \label{tab1}
\begin{tabular}{*{4}{c}}
\hline 
$\gamma_{min}$ & $U_e/U_{th}$ & $\dot{ \epsilon}_h$ & $\dot{ \epsilon}_h/\dot{ \epsilon}_c$ \\
 & & erg cm$^{-3}$ s$^{-1}$ & \\
\hline 
3 & $7.7\%$ & $1.1\times10^{-26}$ & 72 \\
5 & $3.3\%$ & $2.9\times10^{-27}$ & 18 \\
10 & $1.2\%$ & $4.5\times10^{-28}$ & 2.8 \\
30 & $0.20\%$ & $2.4\times10^{-29}$ & 0.15 \\
100 & $0.026\%$ & $1.0\times10^{-30}$ & $6.3\times10^{-3}$ \\
200 & $0.0073\%$ & $1.6\times10^{-31}$ & $9.9\times10^{-4}$ \\
 \hline
 \end{tabular}
 \end{center}
 \end{table}

\begin{figure}[t]
\includegraphics[width=\columnwidth]{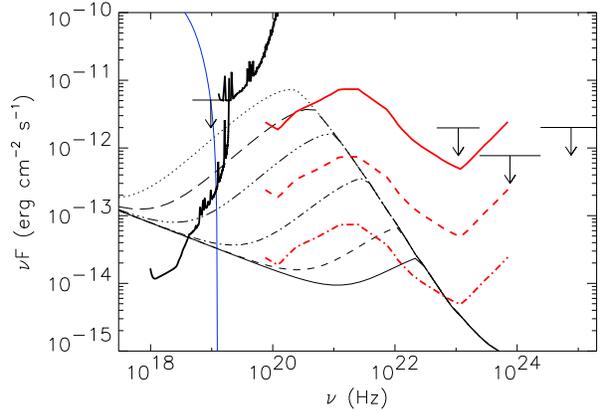}
\caption{High energy emission in the Coma cluster produced via ICS and NTB by non-thermal electrons with spectrum as in Eq.(\ref{spettroele}) with $s_e=3.7$ and different values of the minimum energy of the electrons: $\gamma_{min}=200$ (solid line), 100 (dashed), 30 (dot-dashed), 10 (three dots-dashed), 5 (long dashed), and 3 (dotted). Plotted are also the upper limits in the hard X-ray band from NuSTAR (Gastaldello et al., 2015) and in the gamma ray from Fermi-LAT (Ackermann et al., 2010), the thermal emission from the Intra Cluster Medium (blue line), the Astro-H (HXI and SGD) sensitivity for 100 ks of time integration (from http://astro-h.isas.jaxa.jp/researchers/sim/sensitivity.html; black thick line), and the sensitivity of e-ASTROGAM for an effective exposure of one year (de Angelis et al., 2018; red solid thick line). For reference also the e-ASTROGAM sensitivity divided by a factor of 10 (red dashed thick line) and 100 (red dot-dashed thick line) is plotted.}
\label{spectrum}
\end{figure}

The high-energy emission provided by relativistic electrons by NTB and ICS for different values of $\gamma_{min}$ is shown in Fig.\ref{spectrum}, where it is compared with the expected sensitivity of e-ASTROGAM for an effective exposure of one year (de Angelis et al., 2018). From this figure, it is possible to see that the expected soft gamma ray emission from diffuse relativistic electrons is below the e-ASTROGAM sensitivity for $\gamma_{min}>5$.

For $\gamma_{min} \le 5$ the soft gamma ray emission is in principle observable with e-ASTROGAM, but the physical conditions in this case are problematic: first, the resulting energy content stored in the relativistic electrons is bigger than 3\%, i.e. of the order of the upper limits on the cosmic ray protons derived from Fermi-LAT upper limits (Huber et al., 2013; Prokhorov \& Churazov, 2014), while it is expected that non-thermal electrons should contain only a fraction of the energy stored in cosmic ray protons of the order of $10^{-2}$ or smaller (e.g. Vazza \& Br{\"u}ggen, 2014). Second, electrons with such a low value of minimum energy should have a very small lifetime, of the order of $10^7-10^8$ yrs because of the Coulomb losses (e.g. Sarazin, 1999), and therefore they should be present only for a small fraction of the cluster lifetime, if no steady sources of acceleration are present. Finally, as it is possible to see from Table \ref{tab1}, where we also show the values of the ratio between the heating and the cooling rates, the heating rate provided by electrons with low $\gamma_{min}$ is much higher than the cooling rate of the gas, and this can be a problem for the stability of the ICM. However, this last problem is not too serious if the electrons have a short lifetime: for example, in the case $\gamma_{min}=3$ the heating rate at the center of the Coma cluster is $\dot{ \epsilon}_h\sim1.1\times10^{-26}$ erg cm$^{-3}$ s$^{-1}$. Assuming the lifetime of these electrons to be of the order of $10^8$ yrs (really it is probably smaller, see Sarazin, 1999), the energy density provided at the center of the cluster during this lifetime is $\sim3.6\times10^{-11}$ erg cm$^{-3}$. This number is smaller than the energy density of the thermal gas at the center of the cluster, $\sim1.3\times10^{-10}$ erg cm$^{-3}$, therefore the heating provided by non-thermal electrons during their lifetime is not sufficient to overheat the ICM. Obviously, this fact excludes the possibility of a steady source for acceleration of electrons, and puts a limit on the time for which NTB from low-energy electrons is in principle observable.

In Fig.\ref{spectrum}, we also show for reference the sensitivity of hypothetical instruments with sensitivities 10 and 100 times better than e-ASTROGAM. In the first case, NTB emission would be detectable for values up to $\gamma_{min}\simgt 10$, and in the second case it would be detectable also for $\gamma_{min}\sim200$, for which the lifetime of non-thermal electrons is expected to be higher than the one of the Universe (Sarazin, 1999). However, such sensitivity levels are much better than the e-ASTROGAM one, and therefore it is hard to believe they can be reached in next years.

From Fig.\ref{spectrum} it is also possible to see that in the Hard X-ray band the NTB can provide a significant contribution compared to the ICS one for small values of $\gamma_{min}$, that in principle can be observed with an instrument like Astro-H, even if in this band this emission is lower than the thermal one, so it would be necessary to disentangle thermal and non-thermal emissions. 

We also calculated the flux in the EUV and Soft X-ray (SXR) bands produced by the non-thermal electrons by ICS of the CMB photons, and compared the results with the flux measured with EUVE in the 0.13--0.18 keV band, $F_{EUV}\sim4.1\times10^{-12}$ erg cm$^{-2}$ s$^{-1}$ (Bowyer et al., 2004), and with the flux measured by ROSAT in the 0.2--0.4 keV band, $F_{SXR}\sim1.1\times10^{-11}$ erg cm$^{-2}$ s$^{-1}$ (Bonamente et al., 2002). The ICS flux calculated in our model, that does not depend on the value of $\gamma_{min}$ because it is produced by electrons with $\gamma>250$, is $F_{ICS}\sim8.0\times10^{-14}$ erg cm$^{-2}$ s$^{-1}$ in the EUV band, and $F_{ICS}\sim1.4\times10^{-13}$ erg cm$^{-2}$ s$^{-1}$ in the SXR band. Both these values are much lower than the observed ones, suggesting that the observed emissions in these bands should have a different origin, like a thermal origin from a low-temperature component of the ICM (e.g. Lieu et al., 1996).

Finally, NTB does not appear to be observable in the Fermi-LAT band, where instead the emission of hadronic origin, not considered in this paper, should be dominant (e.g. Colafrancesco \& Blasi, 1998).

\section{Discussion and conclusions}

The study of the low-energy part of the spectrum of relativistic electrons in galaxy clusters by detecting their NTB appears to be a hard task with the next generation of instruments operating in the soft gamma ray band.

As pointed out by Petrosian (2001), NTB is inefficient compared with the Coulomb losses rate: the consequence is that the electrons with low energy have a very short lifetime compared to the cluster one, and that the heating rate provided by these electrons is very high, while the NTB emission is not very strong. 

We have found that in principle there is a narrow region of the parameters space (with $\gamma_{min}\le 5$) where the NTB in the Coma cluster can be observed with e-ASTROGAM. However, these electrons can be present only for a short fraction of the cluster lifetime because of their short lifetime, and because otherwise they would overheat the cluster gas. Also, the energy content of these electrons should be of the order of the upper limits on cosmic ray protons energy content in galaxy clusters: this can be a problem because, on the basis of our present understanding of mechanisms of particles acceleration in galaxy clusters, one should expect that the energy content of the electrons should be lower than the one of protons (e.g. Vazza \& Br\"uggen, 2014), even if some attempt to identify some acceleration mechanism efficient on electrons but not on protons in galaxy clusters has been done (e.g. Wittor, Vazza \& Br\"uggen, 2016).

On the other hand, since several studies of non-thermal SZE in galaxy clusters (Colafrancesco et al., 2011, 2014; Marchegiani \& Colafrancesco, 2017) indicate a possible low value of $\gamma_{min}$ and a high energy content stored in non-thermal electrons, it still appears useful to try to detect galaxy clusters with e-ASTROGAM in order to check these results.

It is also necessary to point out that, even if a cluster would be in principle observable, other issues could be present in order to establish if the observed emission would be really produced by diffuse electrons. In fact, the expected angular resolution of e-ASTROGAM, although improved compared to previous gamma ray instruments, in the energy range 30--100 MeV is of the order of 1.5 degrees (de Angelis et al., 2018), therefore basically all cluster core regions (including Coma) should appear as point-like sources; therefore, the presence of other sources with a soft gamma ray emission (e.g. AGNs) in the field of view of the instrument would be difficult to remove from the cluster one. 

Another possible problem can be the gamma ray emission produced by decay of neutral pions produced in hadronic interactions between cosmic ray protons and thermal gas nuclei (e.g. Colafrancesco \& Blasi, 1998). Although still not firmly detected in galaxy clusters (e.g. Ackermann et al., 2016), this emission should be dominant at energies of the order of 70 MeV or more, and this is another factor that limits the possibility to detect the NTB in galaxy clusters only to the case of low values of $\gamma_{min}$.

In order to improve the situation, it would be necessary an instrument with a sensitivity at least 10 times (or preferably more) better than e-ASTROGAM. However, considering that this instrument adopts very advanced technological solutions for the present day, it is difficult to think that these improved sensitivity levels can be reached in next years. We also note that, since most of other rich clusters are more distant than Coma, the situation is not expected to be easier in other clusters, especially in clusters that do not host a radio halo, suggesting that the non-thermal activity in these clusters can be lower than in the Coma case.

\section*{Acknowledgments}
This work is based on the research supported by the South African Research Chairs Initiative
of the Department of Science and Technology and National Research Foundation of South
Africa (Grant No 77948). PM  acknowledges support from the Department of Science and Technology/National Research Foundation
(DST/NRF) Square Kilometre Array (SKA) post-graduate bursary initiative under the same Grant.
I thank the Referee for useful comments and suggestions.

\end{document}